\documentclass[a4paper,11pt]{article}
\usepackage{pos}

\title{Exploring Earth's Matter Effect in High-Precision Long-Baseline Experiments }

\author*[a,b]{Masoom Singh}
\author[b,c,d]{Sanjib Kumar Agarwalla}

\affiliation[a]{Utkal University, Vani Vihar, Bhubaneswar, Odisha 751004, India,}

\affiliation[b]{Institute of Physics, Sachivalaya Marg,  Sainik School Post, Bhubaneswar 751005, India}

\affiliation[c]{Homi Bhabha National Institute, Training School Complex, Mumbai 400094, India}

\affiliation[d]{International Centre for Theoretical Physics, Strada Costiera 11, Trieste 34151, Italy}

\emailAdd{masoom@iopb.res.in (ORCID: 0000-0002-8363-7693)}
\emailAdd{sanjib@iopb.res.in (ORCID: 0000-0002-9714-8866)}

\abstract{ 
The Earth's matter effect is going to play a crucial role in measuring the unknown three-flavor neutrino oscillation parameters at high confidence level in future high-precision long-baseline experiments. We observe that owing to the new degeneracies among the most uncertain oscillation parameters $\left(\delta_{\mathrm{CP}},~\theta_{23}\right)$ and the average Earth's matter density $\left(\rho_{\mathrm{avg}}\right)$ for the 1300 km baseline, the sensitivity of the upcoming Deep Underground Neutrino Experiment (DUNE) to establish Earth's matter effect reaches only about 2$\sigma$ C.L. for all possible choices of oscillation parameters. We notice that the current uncertainty in $\delta_{\mathrm{CP}}$ degrades the measurement of $\rho_{\mathrm{avg}}$ more as compared to $\theta_{23}$. To lift these degeneracies, we explore the possible complementarity between DUNE and Tokai to Hyper-Kamiokande (T2HK/JD) facility with a second detector in Korea, popularly known as T2HKK or JD+KD setup. While DUNE uses wide-band beam with on-axis detector, T2HKK setup plans to use narrow-band beam with two off-axis detectors: one in Japan and other in Korea. We exhibit how the high-precision measurement of $\delta_{\mathrm{CP}}$ in JD+KD setup and the information on $\rho_{\mathrm{avg}}$ coming from DUNE can reduce the impact of these degeneracies in both $(\rho_{\mathrm{avg}}-\delta_{\mathrm{CP}})$ and $(\rho_{\mathrm{avg}}-\theta_{23})$ planes. We show that the combined data from DUNE and JD+KD setups can establish Earth’s matter effect at more than 6$\sigma$ C.L. irrespective of both the choices of mass hierarchy: normal (NH) and inverted (IH), $\delta_{\mathrm{CP}}$, and $\theta_{23}$. With the help of this combined data set, we can measure the average matter density $\left(\rho_{\mathrm{avg}}\right)$ with a relative 1$\sigma$ precision of around 11.2\% (9.4\%) assuming true NH (IH) and $\delta_{\mathrm{CP}} = -90^{\circ}/90^{\circ}$.
}

\FullConference{%
  *** The European Physical Society Conference on High Energy Physics (EPS-HEP2021), ***\\
  *** 26-30 July 2021 ***\\
  *** Online conference, jointly organized by Universität Hamburg and the research center DESY ***
}

\begin{document}
\maketitle
\vspace{-0.7cm}
\section{Complementarity between DUNE and T2HKK (JD+KD) Setups}
\vspace{-0.2cm}
In this work, we explore the interesting complementarity between the two next generation high-precision long-baseline experiments DUNE and T2HKK (JD+KD) in establishing the Earth's matter effect \cite{MSW} by rejecting the vacuum oscillation. The DUNE far detector (a 40 kt LArTPC) will receive an on-axis, high-intensity, wide-band neutrino beam covering both first and second oscillation maxima with a baseline of 1300 km \cite{DUNE}. On the other hand, the T2HKK setup plans to house its first far detector (187 kt, water Cherenkov detector) in Japan (JD) at a distance of 295 km from J-PARC and to deploy an another 187 kt, water Cherenkov detector in Korea (JD+KD) at a baseline of 1100 km \cite{T2HKK}. The Japanese (Korean) detector will observe an off-axis (2.5$^{\circ}$), narrow-band beam covering first (second) oscillation maximum. We expect a high-precision measurement of $\delta_{\mathrm{CP}}$ and a conclusive evidence for leptonic CP violation from this JD+KD setup, which has very less matter effect. On the other hand, DUNE feels substantial matter effect due to its larger baseline and energies as compared to the JD+KD setup. Therefore, the combined data from DUNE and JD+KD may establish the Earth's matter effect at high C.L. by reducing the impact of possible degeneracies among the oscillation parameters $\left(\delta_{\mathrm{CP}},~\theta_{23}\right)$ and the average Earth's matter density $\left(\rho_{\mathrm{avg}}\right)$. In this work, we discuss several interesting issues along this direction. 
\vspace{-0.25cm}
\section{Establishing Earth's matter effect}
\vspace{-0.2cm}
We perform our simulations using the GLoBES software \cite{globes}. We generate the prospective data with the following choices of oscillation parameters: $\sin^{2}\theta_{23} = [0.44\, , 0.5\, , 0.56], ~\sin^2 2\theta_{13}= 0.085$, $\sin^2\theta_{12}=0.307$, $\delta_{\mathrm{CP}}$ in the range -180$^{\circ}$ to 180$^{\circ}$, $\Delta m^2_{31}=2.5 (-2.4)\times 10^{-3}\,\mathrm{eV}^2$ for NH (IH), $\Delta m^2_{21}=7.4 \times 10^{-5}\,\mathrm{eV}^2$, and the average matter density $(\rho_{\mathrm{avg}}) = 2.86$ g/cm$^{3}$ for all the three (JD, KD, and DUNE) baselines.
\begin{figure}[ht!]
\centering
   \includegraphics[width=0.84\textwidth,height=6cm]{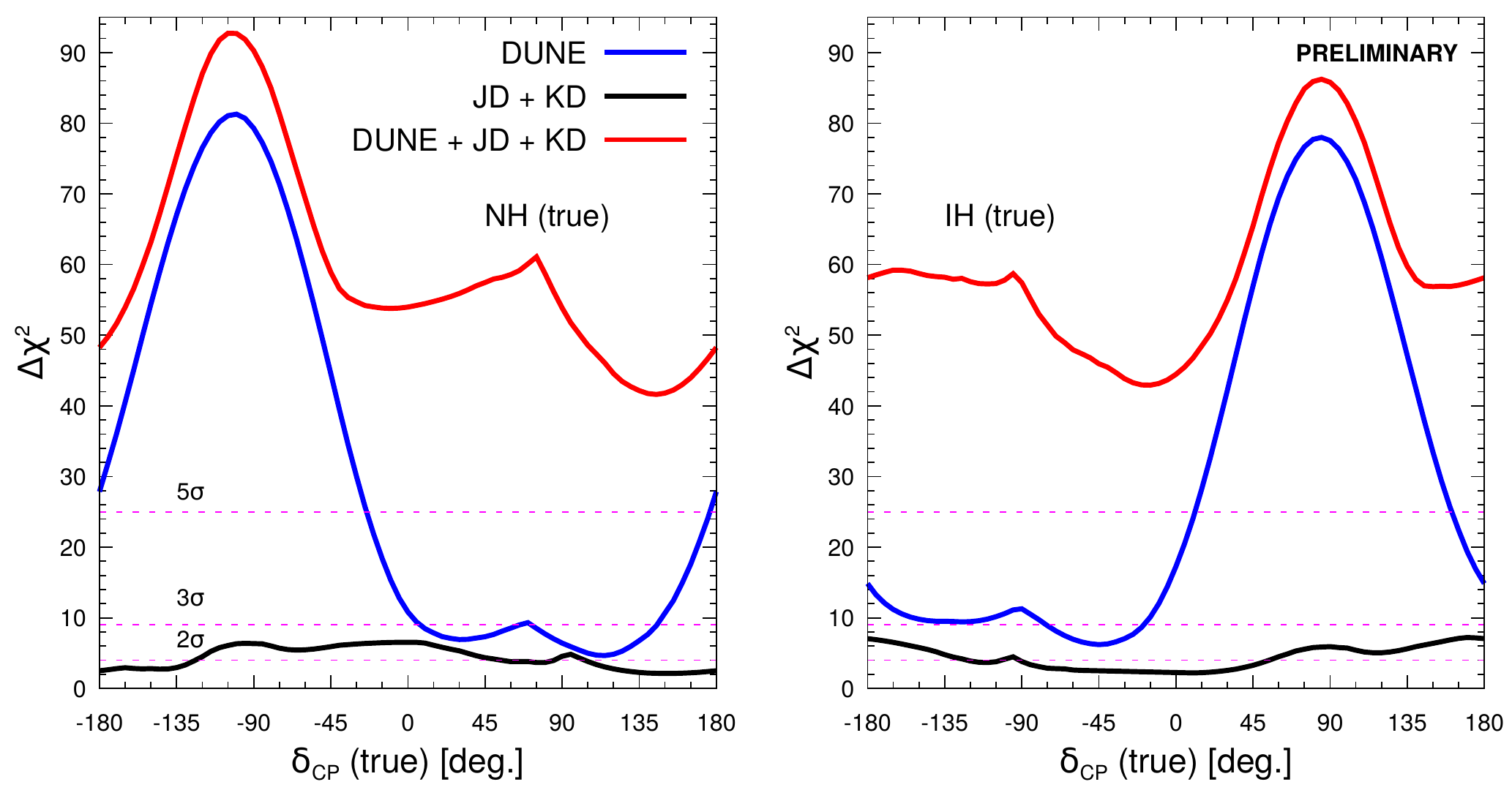}
 \caption{\footnotesize{The sensitivity of JD+KD (black  curves), DUNE (blue curves), and the combined DUNE+JD+KD setup (red curves) in establishing the Earth's matter effect as a function of true $\delta_{\mathrm{CP}}$ assuming true NH (IH) in the left (right) panel. We consider $\sin^{2}\theta_{23}$ (true) = 0.5 and marginalize over $\sin^{2}\theta_{23} = [0.4 : 0.6]$ , $\delta_{\mathrm{CP}}= [-180^{\circ}:180^{\circ}]$ , and $\Delta m^2_{31}= \pm \,[2.36 : 2.64]\times 10^{-3}$ in the fit.}}
 \label{fig:2}
 \end{figure}
The statistical significance of the long-baseline experiments to establish the Earth's matter effect by refuting the vacuum oscillation is defined as follows
\begin{equation}
\Delta \chi^2 = \underset{(\vec{\gamma},~ \lambda_{\mathrm{1}},~ \lambda_{\mathrm{2}})}{\mathrm{min}} \{ \chi^2(\rho_{\mathrm{avg}}^{\mathrm{true}} \neq 0) - \chi^2(\rho_{\mathrm{avg}}^{\mathrm{test}} = 0) \}~, 
\label{eq1}
\end{equation}

\vspace{-0.8cm}
where $\vec{\gamma}=\{\theta_{23},~\delta_{\mathrm{CP}},~ \Delta m_{31}^2\}$ is the set of oscillation parameters on which marginalization is performed and $\lambda_{1}$, $\lambda_{2}$ are the systematic pulls \cite{pull} on signal and background, respectively. In Fig.\ref{fig:2} we observe that DUNE (blue lines) itself can establish the matter effect for about 45\% choices of true $\delta_{\mathrm{CP}}$ at 5$\sigma$ C.L. for both true NH and IH. On the other hand, the JD+KD setup (black lines) alone has very less sensitivity towards the Earth's matter effect. When we combine the performance of DUNE and JD+KD, we observe a significant enhancement in the sensitivity and Earth's matter effect can be established with more than 6$\sigma$ C.L. (red lines) for all possible choices of true $\delta_{\mathrm{CP}}$ and for both NH and IH. We see this improvement in the sensitivity for the unfavorable choices of true $\delta_{\mathrm{CP}}$ (around $0^{\circ}$ to $180^{\circ}$ for true NH and $-180^{\circ}$ to $0^{\circ}$ for true IH)  because the data from JD+KD setup reduces the impact of marginalization over test $\delta_{\mathrm{CP}}$ while analyzing the data from DUNE. 
\vspace{-0.4cm}
\section{Precision measurement of $\mathbf{\rho_{\mathrm{\textbf{avg}}}}$}
\vspace{-0.2cm}
The statistical significance to measure $\rho_{\mathrm{avg}}$ in a given experiment is defined as
\vspace{-0.2cm}
\begin{equation}
\Delta\chi^{2}_{\mathrm{PM}} (\rho_{\mathrm{avg}}) = \chi^{2}(\rho_{\mathrm{avg}}) - \chi^{2}_{0}~,
\end{equation}
where we obtain $\chi^{2}\left(\rho_{\mathrm{avg}}\right)$ by performing a fit to the prospective data with $\rho_{\mathrm{avg}} = 2.86$ g/cm$^{3}$ and $\chi_{0}^{2}$ is the minimum value of $\chi^{2}(\rho_{\mathrm{avg}})$ considering $\rho_{\mathrm{avg}}$ in the range of 1.5 to 4 g/cm$^{3}$.
\vspace{-0.2cm}
\begin{figure}[hbt!]
\centering
\includegraphics[width=0.74\textwidth,height=6cm]{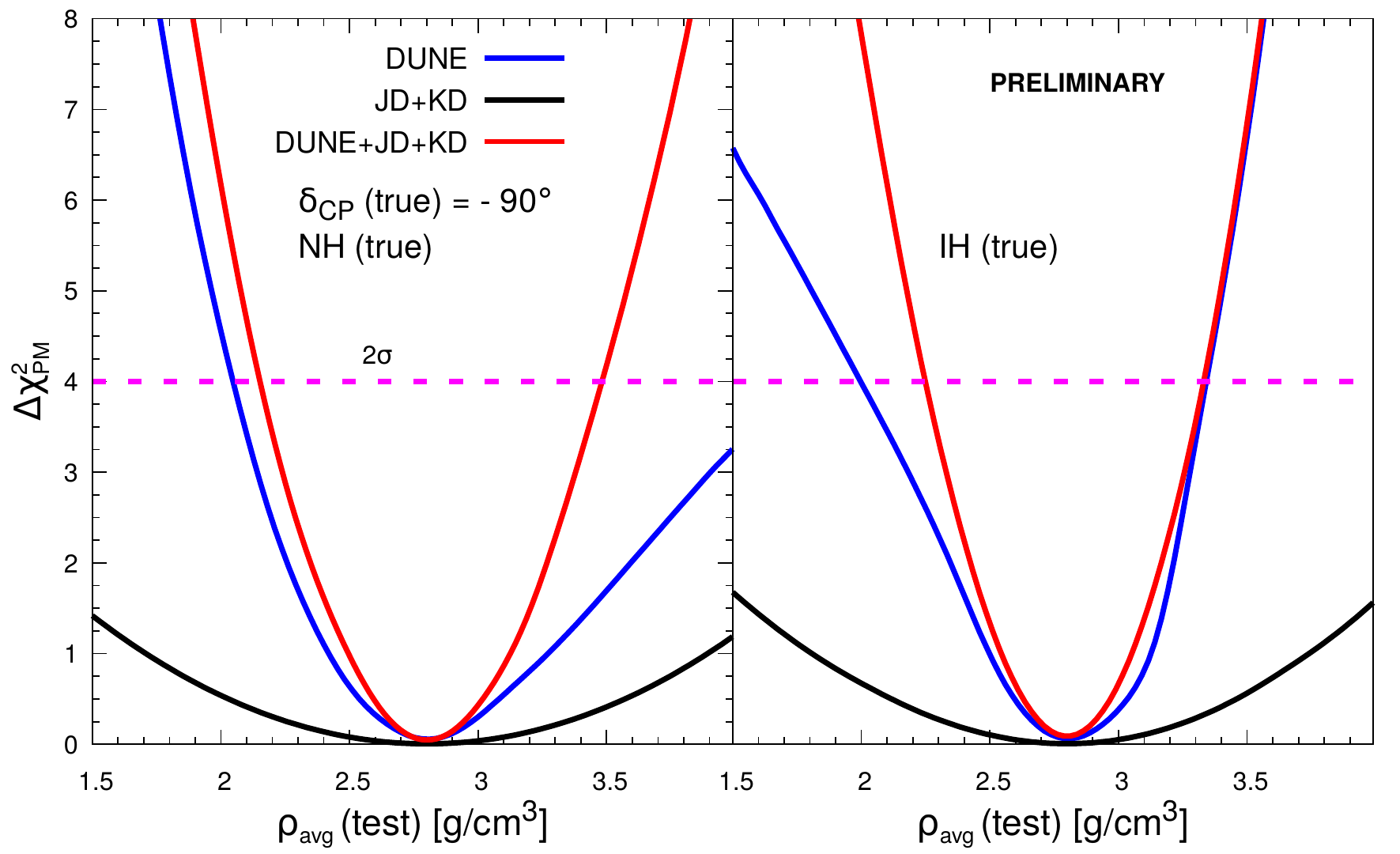}
\caption{Left (Right) panel shows the achievable precision in the measurement of $\rho_{\mathrm{avg}}$ for  JD + KD (black lines), DUNE (blue lines), and DUNE+JD+KD (red lines) assuming true NH (IH). Here, we consider true $\delta_{\mathrm{CP}}$ = -90$^{\circ}$ and $\sin^{2}\theta_{23}$ = 0.5. In the fit, we marginalize over $\sin^{2}\theta_{23},~\delta_{\mathrm{CP}},$ and $\Delta m^{2}_{31}$ (see Fig. \ref{fig:2} caption).}
\label{fig2} 
\end{figure}
Fig. \ref{fig2} shows that the JD+KD setup alone offers a relative 1$\sigma$ precision in $\rho_{\mathrm{avg}}$ of around 40\% (35\%) for true NH (IH) assuming $\delta_{\mathrm{CP}}$ (true) = -90$^{\circ}$ and $\sin^{2}\theta_{23}$ (true) = 0.5. The same for DUNE setup alone is around 15\% (12\%). Interestingly, when we combine the data from these two high-precision experiments, the achievable precision in $\rho_{\mathrm{avg}}$ reaches to 11.2\% (9.4\%).
\section{Degeneracies in test $\left(\rho_{\mathrm{avg}} - \delta_{\mathrm{CP}}\right)$ and test $\left(\rho_{\mathrm{avg}} - \theta_{23}\right)$ Planes}
\begin{figure}[hbt!]
\begin{minipage}{0.446\textwidth}
\centering
\includegraphics[width=1.0\linewidth,height=1.07\linewidth]{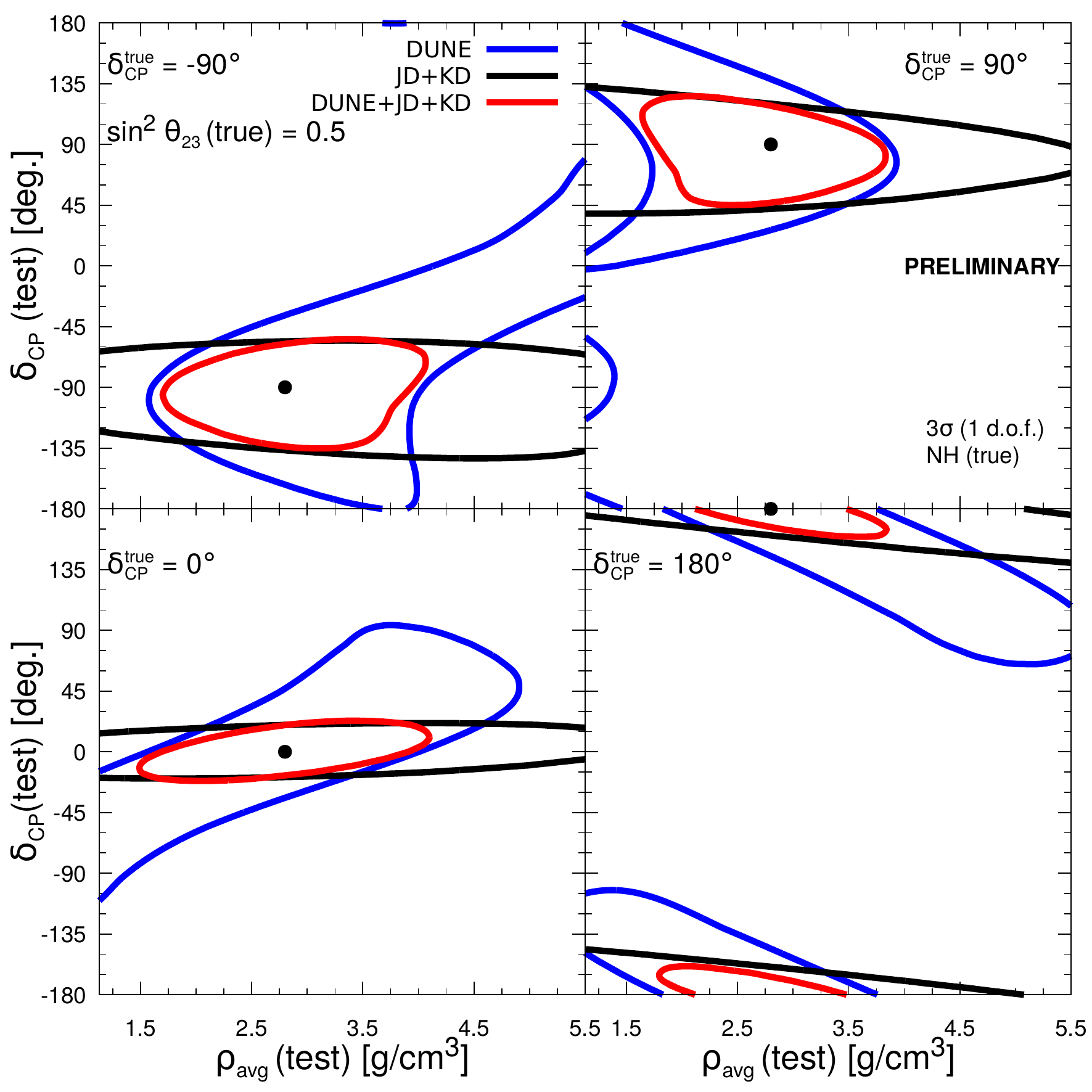}
\end{minipage}\hfill
\begin{minipage}{0.55\textwidth}
\centering
\includegraphics[width=1.03\linewidth , height=0.44\linewidth]{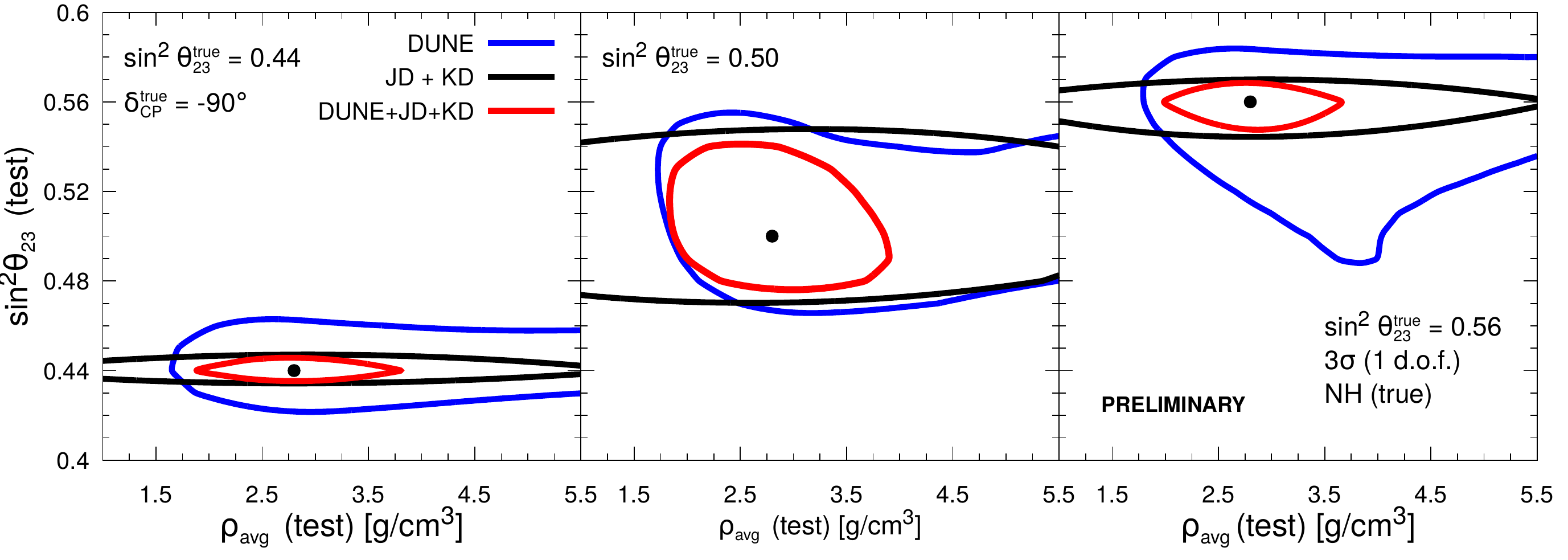}
\caption{\footnotesize{Left panel shows the allowed regions in test $(\rho_{\mathrm{avg}}-\delta_{\mathrm{CP}})$ plane for four true values of $\delta_{\mathrm{CP}} = -~90^{\circ}, 90^{\circ}, 0^{\circ},~\text{and}~ 180^{\circ}$ assuming $\sin^{2}\theta_{23}$ (true) = 0.5. Right panel depicts the same in test $(\rho_{\mathrm{avg}}-\theta_{23})$ plane for three true choices of $\sin^{2}\theta_{23}= 0.46,~ 0.5,~ \text{and}~0.56$ for $\delta_{\mathrm{CP}}$ (true) =  $-~90^{\circ}$. The black, blue, and red curves are for JD + KD, DUNE, and DUNE + JD + KD, respectively at 3$\sigma$ C.L. (1 d.o.f.) assuming true NH.}}
\label{fig4} 
\end{minipage}
\end{figure}

The black curves in left (right) panel of Fig. \ref{fig4} shows that the JD+KD setup alone can measure $\delta_{\mathrm{CP}} ~ \left(\theta_{23}\right)$ quite precisely while having almost no sensitivity towards $\rho_{\mathrm{avg}}$ due to their shorter 

baselines. Whereas the DUNE setup alone can constrain the allowed ranges in $\rho_{\mathrm{avg}}$ and can provide reasonable measurements of $\delta_{\mathrm{CP}}$ and $\theta_{23}$ (blue curves). When we combine the data from these two setups, we see a considerable reduction in the allowed ranges in both $\left(\rho_{\mathrm{avg}} - \delta_{\mathrm{CP}}\right)$ and $\left(\rho_{\mathrm{avg}} - \theta_{23}\right)$ planes (red curves) due to the complementary information from these two experiments.
\vspace{-0.45cm}
\section{Conclusion}
\vspace{-0.1cm}
DUNE with 1300 km baseline has significant matter effect and can measure $\delta_{\mathrm{CP}}$ and $\theta_{23}$ with reasonable precision exploiting the information on oscillation pattern at several $L/E$ values. On the other hand, with a relatively shorter baseline and high statistics JD offers an unmatched sensitivity to the $\delta_{\mathrm{CP}}$ free from matter effect. KD with a roughly four times baseline than JD has some sensitivity to Earth's matter effect and provides crucial information on $\delta_{\mathrm{CP}}$ around the second oscillation maxima. In this work, for the first time, we show how the complementary features between DUNE and JD+KD setups can play an important role to establish the Earth's matter effect at more than 6$\sigma$ C.L. for any values of oscillation parameters. The complementary informations coming from DUNE and JD+KD setups also play an important role to provide a high-precision measurement of $\rho_{\mathrm{avg}}$ and to reduce the allowed regions in $\left(\rho_{\mathrm{avg}} - \delta_{\mathrm{CP}}\right)$ and $\left(\rho_{\mathrm{avg}} - \theta_{23}\right)$ planes considerably.

\textbf{Acknowledgements :} M.S. acknowledges financial support from DST, Govt. of India (DST/INSPIRE Fellowship/2018/IF180059). S.K.A. acknowledges financial support from DAE, DST, DST-SERB, Govt. of India, and INSA. The numerical simulations are performed using SAMKHYA: High-Performance Computing Facility at Institute of Physics, Bhubaneswar.

\end{document}